**Title:**
Oxidation Kinetics of Superconducting Niobium and $\alpha$-Tantalum in Atmosphere at Short and Intermediate Time Scales

**Authors:**
Hunter J. Frost[1,2], Ekta Bhatia[3], Zhihao Xiao[3], Stephen Olson[3], Corbet Johnson[3], Kevin Musick[3], Thomas Murray[3], Christopher Borst[1,3], Satyavolu Papa Rao[3]

**Affiliations:**
[1]College of Nanotechnology, Science, and Engineering, University at Albany, Albany, NY, USA
[2]TEL Technology Center, America, LLC., Albany, NY, USA
[3]NY CREATES, Albany, NY, USA

**Abstract:**

The integration of superconducting niobium and tantalum into superconducting quantum devices has been increasingly explored over the past few years. Recent developments have shown that two-level-systems (TLS) in the surface oxides of these superconducting films are a leading source of decoherence in quantum circuits, and understanding the surface oxidation kinetics of these materials is key to enabling scalability of these technologies. We analyze the nature of atmospheric oxidation of both niobium and $\alpha$-tantalum surfaces at time scales relevant to fabrication, from sub-minute to two-week atmospheric exposure, employing a combination of x-ray photoelectron spectroscopy and transmission electron microscopy to monitor the growth of the surface oxides. The oxidation kinetics are modeled according to the Cabrera-Mott model of surface oxidation, and the model growth parameters are reported for both films. Our results indicate that niobium surface oxidation follows a consistent regime of inverse logarithmic growth for the entire time scale of the study, whereas $\alpha$-Ta surface oxidation shows a clear transition between two inverse logarithmic growth regimes at time t ≈ 1 hour, associated with the re-coordination of the surface oxide as determined by x-ray photoelectron spectroscopy analysis. Our findings provide a more complete understanding of the differences in atmospheric surface oxidation between Nb and $\alpha$-Ta, particularly at short time scales, paving the way for the development of more robust fabrication control for quantum computing architectures.

**Introduction:**

In recent decades, many significant advances have been made toward the fabrication of large-scale quantum computing architectures. Specifically, superconducting quantum circuitry has emerged as a promising hardware implementation to solve complex problems considered impossible for even the most efficient classical computing platforms (1−4). However, scaling this technology requires continued progress to improve reliability and performance, specifically relating to the fabrication of high-quality materials and an increased understanding and control over defects in interfaces and surfaces (5).

Historically, the superconducting material of choice for qubit and Josephson junction fabrication has been aluminum (6-10). The relatively wide-spread use of aluminum has a few explanations: 1) aluminum has a critical temperature ($T_c$) of about 1.2 K, which is sufficiently high for quantum computing systems that operate at milliKelvin (mK) temperatures, 2) aluminum can be deposited using techniques like sputtering and thermal evaporation, under conditions where an organic resist bridge can be safely used for 'shadow' evaporation, 3) aluminum naturally forms a thin oxide layer ($Al_2O_3$) whose thickness can be controlled by exposure time to low pressures (~100 mTorr) of oxygen, in situ in the deposition system (11-13), and 4) the early development of superconducting qubits focused on aluminum, leading to significant research and optimization, and as a result, a relatively comprehensive body of knowledge and established techniques for using aluminum in quantum applications (6-13).

For all superconducting metals used as electrodes for Josephson junctions, coplanar waveguide resonators, and other wiring of quantum computing circuits, considerable literature has accumulated showing that surface oxides that form upon atmospheric exposure serve as major sources of microwave dissipation (14). At standard mK operating temperatures, they can display loss tangent values that are 3 orders of magnitude larger than the superconducting electrodes and device substrates (15−17). This loss can be largely attributed to two-level system (TLS) defects in the amorphous surface oxide (15). The system can then transition between the two states through low energy excitations, leading to the decoherence of an otherwise coherent quantum state (18,19). Though an understanding of the fundamental mechanism for TLS decoherence remains unclear (whether due to atomic reconfiguration (20), tunneling of electrons (21), paramagnetic spins (22), quasiparticle traps (23), or coupling to phonons (24)), it is clear that a reduction in the thickness of surface oxides and an understanding of their growth in ambient conditions is critical to minimizing TLS impacts in large-scale quantum devices (17).

Both niobium (25-28) and tantalum (29-34) have been integrated in quantum circuitry, though the early stages of formation of their surface oxides are less well documented. The surface oxidation kinetics of niobium have been reported in the literature, though these investigations have primarily focused on oxidation at high temperatures, often from 773 – 1573 K and oxidation at long time scales, on the order of days to months (35-40). The surface oxidation kinetics of tantalum have also been reported in the literature, though similarly, these investigations focus on oxidation at high temperatures from 573 – 2073 K and oxidation at long time scales, on the order of days to months (41-47). A comprehensive study of the oxidation kinetics of niobium and tantalum films at room temperature and atmospheric pressure with a primary focus on short to intermediate time scales has yet to be reported. This knowledge would be exceedingly valuable in the interfacial control of these metals when used in fabricating superconducting quantum circuitry.

We employ a combination of x-ray photoelectron spectroscopy (XPS) and transmission electron microscopy (TEM) to investigate the atmospheric oxidation kinetics of superconducting niobium and $\alpha$-tantalum in detail. We monitor the growth rate, atomic composition, and relative oxidation states of both metals. The use of surface-sensitive techniques to examine the superconductor surface also allows us to probe the presence of sub-stoichiometric oxides on the metal's surface, and elucidate the relationships between oxidation states, surface oxidation rate, and surface oxide thickness. This can provide a more complete understanding necessary for decreasing decoherence in quantum computing devices.

**Methodology:**

All experiments were conducted with sufficiently thick Nb and $\alpha$-Ta films (100 nm and 70 nm, respectively) deposited by DC magnetron sputtering in ultra-high vacuum (UHV, base pressure <$10^7$ Torr) using state-of-the-art 300 mm thin film deposition platforms. All experiments were deposited on 780 μm thick 300 mm Si (100) wafers. The crystallinity of the deposited films is reported in Figure 1.

To determine the effects of atmospheric exposure for the Nb and $\alpha$-Ta films, a cyclic series of XPS measurements were performed using an industry-standard inline XPS/XRF system capable of handling 300 mm wafers, which uses normal take-off angle and an Al-K$\alpha$ x-ray with an energy of 1486.6 eV. To evaluate surface oxidation, particular emphasis was given to O1s, Nb3d, and Ta4f binding energy windows (520-540 eV, 190-220 eV, and 12-42 eV, respectively). XPS spectra were collected at three radial locations on the wafer (center, mid-radius, and wafer edge, at r = 0mm, 72.5mm, and 145mm, respectively), and the sample was returned to atmosphere after spectra collection was completed. The sample was intentionally exposed to "fab ambient", actively controlled for temperature at 20.5C ± 1.0C and relative humidity at 30% ± 5%, for a documented length of time before being re-introduced to UHV on the XPS platform for a subsequent measurement. This process was repeated between 12-20 times for each sample, with a particular emphasis placed on the initial 5-200 minutes of atmospheric exposure. Monitoring the atmospheric exposure time of the films was documented with Coordinated Universal Time (UTC) as synchronized to NIST/USNO Official U.S. Time (48). Following a final measurement, select samples were capped with a queue-time minimized PVD-CuMn deposition to cut off atmospheric exposure to the native oxide. Lamellas were prepared from these samples to calibrate surface oxide thickness with TEM imaging, using a 30 kV focused Ga+ ion beam to mill the lamellas. The samples were finely polished to a thickness of roughly 100 nm using 5 kV and 2 kV Ga+ ions in an effort to remove surface damage and amorphization from the regions of interest.

All XPS spectra were analyzed with CasaXPS(49) (version 2.3.19PR1.0). Spectral peak positions were calibrated with $Ta^{+0}$ and $Nb^{+0}$ peak positions, respectively, to account for any surface charge accumulation. To quantify the relative abundance of oxide species, a two-component system with Tougaard background subtraction was utilized for the O1s spectral region. The primary O1s spectral component for both Ta and Nb is fit with a pure Gaussian and accounts for approximately 80% of the total O1s spectral area. Another major O1s peak, fit with an asymmetric Lorentzian and encompassing approximately 20% of the total O1s spectral area, has been added to account for the higher binding energy shoulder. This is proposed to be due to defective sites within the oxide crystal (51-53), adsorbed oxygen (54), or hydroxide species (55). Ta4f and Nb3d spectral regions were fit with Shirley background subtraction, and their spectral component constraints are detailed in Table 1. Spin-orbit coupling constraints were derived from the Phi Electronics Handbook for XPS (56) and the presence and position of sub-oxides compared to relevant references for Nb (25, 57) and Ta (58-60).

We fit the full dataset of XPS spectra for each atmospheric exposure series simultaneously by applying relevant constraints to the first data point and propagating these constraints to the remainder of the data series. Fitting all spectra at once constrains the relative peak positions, relative peak intensities, peak widths, line-shapes, and skewnesses, which reduces the number of free fit parameters and increases our confidence in the fitted photoelectron intensities (61, 62). Following data fitting, relative component areas were extracted for further analysis.

| Photoelectron Line | Quantification Region | Background | Spin-Orbit Coupling Doublet Spacing (eV) | Doublet Area Constraint (%) | Sub-Set | Position Constraint | | | | | | | |
|---|---|---|---|---|---|---|---|---|---|---|---|---|---|
| --- | --- | --- | --- | --- | --- | Comp #1 | Comp #2 | +0 | +1 | +2 | +3 | +4 | +5 |
| O1s | 528-540 eV | Tougaard | N/A | N/A | BE | 528-540 eV | +0.96 | N/A | N/A | N/A | N/A | N/A | N/A |
|  |  |  |  |  | LS | SGL(0) | LA(1.3,20,50) | N/A | N/A | N/A | N/A | N/A | N/A |
| Nb3d | 200-216 eV | Shirley | 2.72 eV | A * 0.667 | BE | N/A | N/A | 200-204 eV | +1.09 | +1.99 | +3.14 | +4.12 | 208-221 eV |
|  |  |  |  |  | LS | N/A | N/A | LA(1.3,20,50) | | | SGL(0) | | |
| Ta4f | 20-33 eV | Shirley | 1.92 eV | A * 0.75 | BE | N/A | N/A | 20-24 eV | +1.10 | N/A | +3.36 | N/A | 27-30 eV |
|  |  |  |  |  | LS | N/A | N/A | LA(1.3,20,50) | | | SGL(0) | | |

Table 1. Constraints for XPS spectral analysis of Nb and $\alpha$-Ta surface oxides, including quantification region (selected to include a sufficient background on either side of the spectral peaks), background type, spin-orbit coupling spacing, doublet area constraint for metal photoelectron lines, as well as the binding energy (BE) constraints and line-shape (LS) selection for each spectral component..

**Results and Discussion:**

*Superconductor Crystal Structure*
In-plane grazing-incidence x-ray diffraction (GIXRD) spectra were collected from Nb and $\alpha$-Ta samples using a 300 mm capable x-ray platform standard in the CMOS IC industry. Analyzed XRD spectra are reported below in Figure 1, indicating polycrystalline body-centered-cubic (bcc) Nb and polycrystalline bcc Ta (known in the literature as $\alpha$-Ta) (63), both of which are attributed to the superconducting phases of their respective materials. The superconducting nature of the metals has been confirmed (manuscripts in preparation).

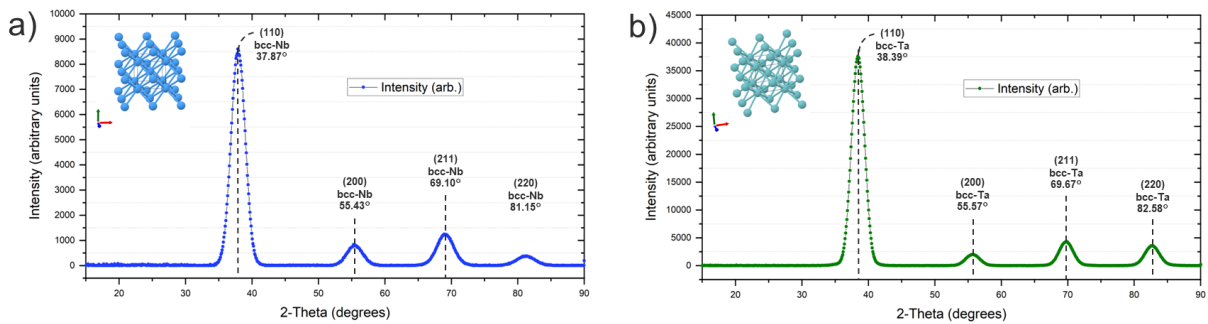

Figure 1. XRD spectra with identified polycrystalline bcc peaks for a) Nb and b) $\alpha$-Ta thin films.

*XPS Spectral Analysis*

To quantify the surface oxide, the following procedure was utilized (see Figure 2). Component areas for O1s and Nb3d/Ta4f were summed to identify the relative XPS intensities for each photoelectron line. Each summation was normalized by a reduced relative sensitivity factor (*r*-RSF) specific to each photoelectron line and sample medium.

The full equation for photoemission intensity generation in a solid is given by:

$$I = nF\sigma\phi yT\lambda \qquad (Eq.\ 1)$$

where I is the number of photoelectrons detected per second from the orbital concerned, n is the concentration of the atom concerned (atoms/cm$^3$), F is the x-ray flux (photons/s/cm$^2$), σ is the partial ionization cross section for the orbital concerned (cm$^2$), y is the fraction of σ retained in the measured peak, ϕ is an angular distribution term, T is the efficiency of detection of the spectrometer (the transmission function), and λ is the inelastic mean free path (IMFP, a function of KE)(61). Making the usual assumptions that F is identical for each point, that the T effect has been corrected for by the tool, that ϕ is unity, and that the measured peaks capture all the signal (i.e. y = 1), then we arrive at the general form:

$$I \approx n\sigma\lambda = n(rRSF) \qquad (Eq.\ 2)$$

This reduced relative sensitivity factor (*r*-RSF) is the default theory equation for converting a relative intensity, $I_a$, into a normalized intensity, $n_a$. This method of quantitative analysis often relies on an assumption of homogeneity in the outermost few nanometers, though in practice, it is quite rare that samples analyzed with XPS meet this criterion, invalidating calculations of concentration when the collection volume is not uniform. However, we propose that our utilization of this equation is not to quantify a concentration of oxygen (or any other element, as is often the case in quantitative XPS analysis), but rather simply to correct the relative intensity by known factors that influence the quantity of detected photoelectrons. For this analysis, photo-ionization cross-sections from each orbital under Al-K$\alpha$ irradiation were extracted from literature data tables (64) and inelastic mean free paths in the given medium were identified from a standard NIST IMFP database (65). All relevant values are reported below in Table 2, including finalized *r*-RSF values for O1s photoelectrons in Ta and Nb, Ta4f photoelectrons in Ta, and Nb3d photoelectrons in Nb.

A further normalization is applied to the O1s spectral intensity to account for sample collection volume variations during analysis, as well as to encompass any potential variation in the assumptions used to reduce Equation 1 to Equation 2, such as small variations in x-ray flux or the efficiency of detection of the spectrometer. The completed normalization equation is as follows:

$$\frac{n_{O1s}}{n_{Met}} = \left(\frac{I_{O1s}}{rRSF_{O1s}} \times \frac{rRSF_{Met}}{I_{Met}}\right) \qquad (Eq.\ 3)$$

where $n_{O1s}$ is the normalized intensity of oxygen, $n_{Met}$ is the normalized intensity of the relevant metal, $I_{O1s}$ and $I_{Met}$ are the relative intensities of the respective elements, and *r*-RSF$_{O1s}$ and *r*-RSF$_{Met}$ are the sensitivity factors as calculated in Table 2. This normalization for O1s intensity by the Nb3d/Ta4f intensity is based in the presumption that any natural variations in collection volume are captured by the relative intensity of the metal photoelectron line.

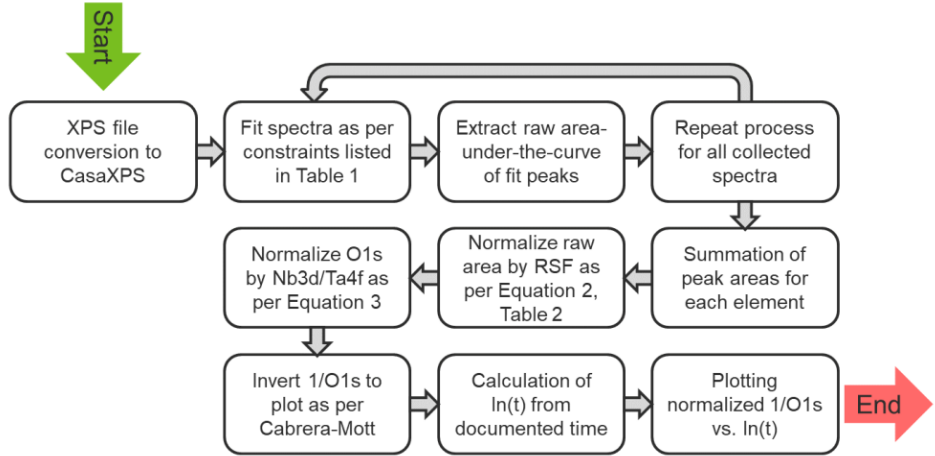

Figure 2. Flowchart for XPS atmospheric oxidation analysis of Nb and Ta surface oxides.

| Photoelectron System | r-RSF Equation | Photo-ionization Cross-Section ($\sigma$) | Inelastic Mean Free Path (IMFP) | Calculated r-RSF |
|---|---|---|---|---|
| O1s in Nb | $\sigma(O1s) * IMFP_{O1s\text{-}in\text{-}Nb}$ | 0.040 | 2.174 | 0.08696 |
| O1s in Ta | $\sigma(O1s) * IMFP_{O1s\text{-}in\text{-}Ta}$ | 0.040 | 1.432 | 0.05728 |
| Nb3d in Nb | $\sigma(Nb3d) * IMFP_{Nb3d\text{-}in\text{-}Nb}$ | 0.1124 | 2.174 | 0.30539 |
| Ta4f in Ta | $\sigma(Ta4f) * IMFP_{Ta4f\text{-}in\text{-}Ta}$ | 0.1274 | 1.957 | 0.24932 |

Table 2. Reduced relative sensitivity factor (r-RSF) calculations for each photoelectron line.

*Cabrera-Mott Oxidation*
The normalized O1s intensity is then plotted according to the Cabrera-Mott model of oxidation to determine the oxidation kinetics of the two superconductors. The Cabrera-Mott model describes the low temperature oxidation of metals through a process involving ionic conduction in the growing oxide layer, where the thickness of the oxide layer increases as oxygen ions migrate inward and react with the underlying metal interface (66). Cabrera and Mott's solution for thin film growth at low temperatures yields an inverse logarithmic growth law (66, 67):

$$\frac{1}{x} = A - B \ln(t) \qquad (Eq.\ 4)$$

where x is the thickness of the surface oxide at a time t, and *A* and *B* are the Cabrera-Mott high-field growth constants. Here, the normalized intensity of the O1s photoelectron line is used as a proxy for surface oxide thickness, and a calibration to thickness with TEM is presented later in this report. To display the data in a linear fashion, the inverse of the normalized O1s intensity and the natural logarithmic function of oxidation time in seconds are the two scales plotted.

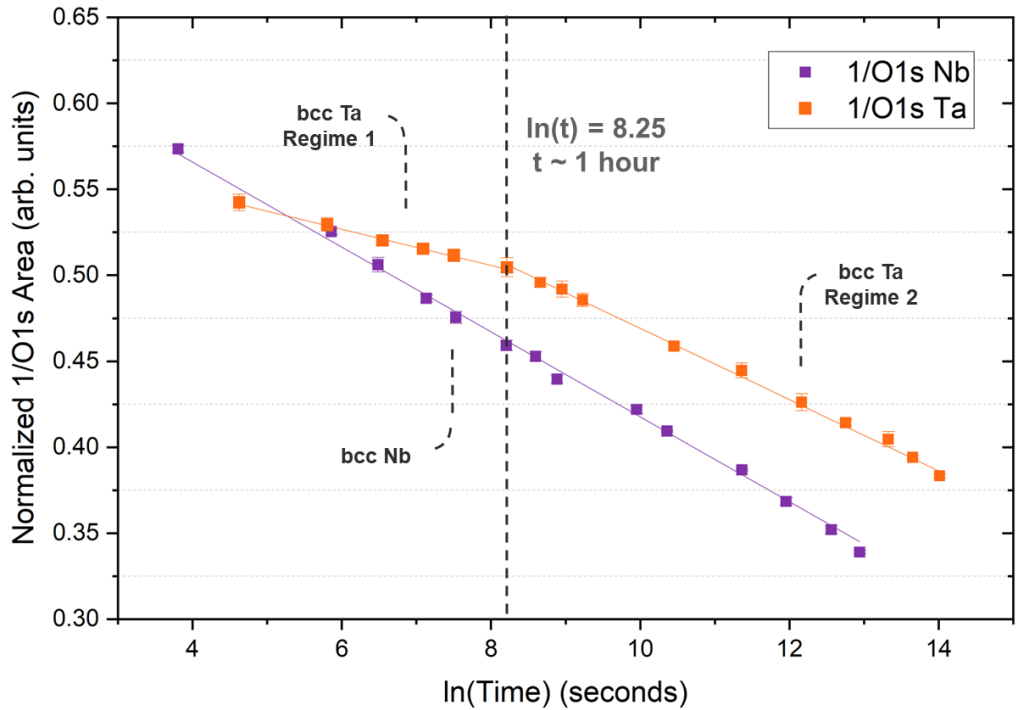

Figure 3. Cabrera-Mott oxidation kinetics of superconducting Nb and $\alpha$-Ta thin films

As can be seen in Figure 3, Nb surface oxide growth provides a good fit (after transforming variables per Equation 4) for the duration of the timescale measured, from 45 seconds (ln(t) = 3.81) to approximately 5 days (ln(t) = 12.94), with Cabrera-Mott growth constants extracted and reported below in Table 3. Conversely, $\alpha$-Ta surface oxide growth exhibits a characteristic break-point after approximately 1 hour of oxidation time (where ln(t) = 8.25), after which the rate of oxide growth increases for the duration of the timescale measured, to approximately 2 weeks of oxidation time (where ln(t) = 14.01). This behavior, in terms of a 'break' in the curve was confirmed in multiple experiments, though only one characteristic set of data is reported above. The Cabrera-Mott growth constants for both regimes are also fit and reported below in Table 3.

| Cabrera-Mott Growth Constant | bcc-Nb | bcc-Ta, Regime 1 | bcc-Ta, Regime 2 |
|---|---|---|---|
| A | 0.6646 | 0.5899 | 0.6765 |
| B | 0.02469 | 0.01053 | 0.02074 |

Table 3. XPS-extracted Cabrera-Mott growth constants from Nb and $\alpha$-Ta atmospheric oxidation

Based on the extracted growth constants, $\alpha$-Ta exhibits a slower rate of oxidation at time scales below 1 hour (Regime 1), then enters a faster oxidation regime (Regime 2). The rate of oxidation for Regime 2 is 1.95 times the oxidation rate of Regime 1. The relative oxidation rates (characterized by the Cabrera-Mott growth constant B) are more similar for Nb and $\alpha$-Ta after the break-point, with Nb oxidizing approximately 19% faster than $\alpha$-Ta. This increased rate of oxidation, coupled with the higher Nb oxidation rate during $\alpha$-Ta Regime 1, contributes to the higher native oxide thicknesses reported in the literature for Nb after long term exposure to air (57).

*TEM Imaging and Calibration to Thickness*
High-magnification TEM images of the surface oxide are reported below in Figure 4, and measurements of the surface oxide were utilized to calibrate the Cabrera-Mott analysis from XPS to thickness. The native oxide thickness was determined by averaging three measurements from independent locations for each sample. Each measurement was determined using relative contrast differences, coupled with the disappearance of underlying $\alpha$-Ta lattice fringing, to define the surface oxide edge.

To calibrate the Cabrera-Mott oxidation curves to thickness, we matched the two points of TEM thickness data to their respective 1/O1s intensities, fit a linear calibration function to the two points, and rescaled the full data-set accordingly in Figure 4. We found the calibration function to be:

$$\frac{1}{Native\ Oxide\ Thickness\ (nm)} = 3.4736 \times \frac{1}{Normalized\ O1s\ Intensity} - 0.988 \quad \text{(Eq. 5)}$$

Following this calibration, the Cabrera-Mott growth constants were re-extracted and are reported below in Table 4.

| Cabrera-Mott Growth Constant | bcc-Ta, Regime 2 XPS O1s | bcc-Ta, Regime 2 TEM Thickness | bcc-Ta, Regime 1 TEM Thickness | bcc-Nb TEM Thickness |
|---|---|---|---|---|
| A | 0.6765 | 1.3619 | 1.0613 | 1.3415 |
| B | 0.02074 | 0.07203 | 0.03559 | 0.08768 |

Table 4. Extracted Cabrera-Mott growth constants after TEM calibration to thickness.

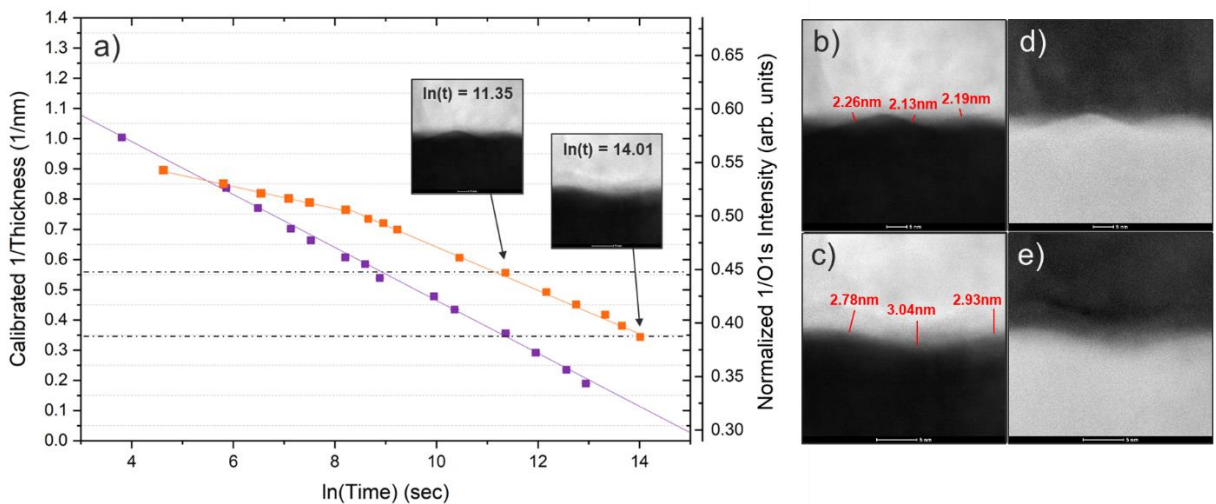

Figure 4. a) Cabrera-Mott oxidation kinetics for Nb and $\alpha$-Ta, calibrated to thickness with b,c) bright-field TEM and d,e) high-angle annular dark field (HAADF) TEM images

*Sub-Stoichiometric Binding Energy Analysis*

To identify the root cause of the $\alpha$-Ta oxidation behavior, the Ta4f and Nb3d binding energy positions for each component in the XPS spectra were extracted. When propagating XPS constraints, the relative positions of the $Ta^{+5}$ and $Nb^{+5}$ oxidation states were not fixed, but rather were constrained to a range of binding energy values, as a variation in the position of these peaks was observed during data collection. An overlay of the collected Ta4f and Nb3d spectra are reported below in Figure 5a & d, and the variations in binding energy of the $Nb^{+5}$ and $Ta^{+5}$ states relative to the $X^{+0}$ states for each respective element are reported as a function of oxidation time $ln(t)$. It may be noted that the onset of binding energy shift observed for $Ta^{+5}$ correlates with the break point observed in the Cabrera-Mott oxidation curves (Figure 5e). Note that Nb has a propensity to form all sub-oxides ($Nb^{+1}$, $Nb^{+2}$, $Nb^{+3}$, and $Nb^{+4}$), whereas a-Ta primarily forms odd sub-oxides ($Ta^{+1}$ and $Ta^{+3}$) (59). At the shortest time scales, the influence of sub-oxide peaks is most prominent, and the overlap of the $Nb^{+4}$ and $Nb^{+5}$ peaks (as seen below in Figure 5c) artificially skews the $Nb^{+5}$ peak position to a lower binding energy. Fitting the other oxidation states of Nb shows a relatively small shift in the $Nb^{+5}$ peak position.. The relatively small shift in the $Nb^{+5}$ peak position is reflected in the lack of any 'break' in the oxidation kinetics of Nb, as shown in Figure 5b.

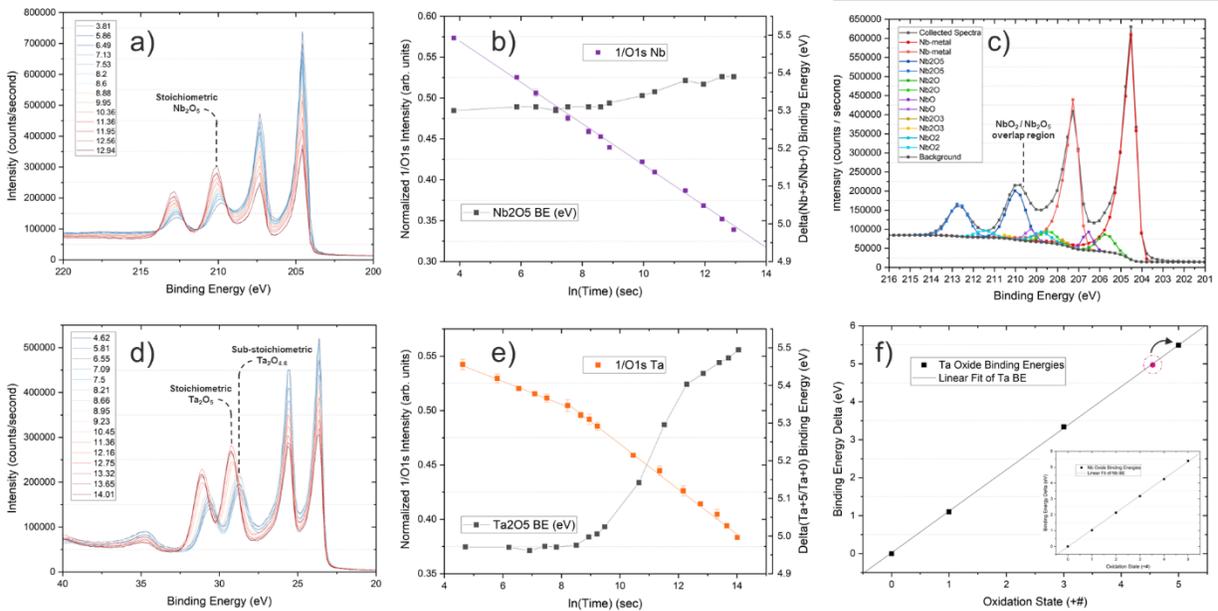

Figure 5. Spectral and sub-stoichiometric binding energy analysis of Nb and $\alpha$-Ta. a) An overlay of all Nb center-point spectra, with b) extracted $Nb_2O_5$ binding energies as a function of oxidation time and c) a characteristic point of Nb3d spectra from $ln(t) = 6.47$, illustrating the peak overlap between the $Nb^{+4}$ and $Nb^{+5}$ oxides. A corresponding overlay and binding energy analysis for $\alpha$-Ta are reported in d,e). A correlation of binding energies and oxidation states for both metals are reported in f).

Based on the relative peak positions, we conclude that in the early stages of oxidation (<1 hour, $ln(t)<8.25$), the $\alpha$-Ta surface initially forms a sub-stoichiometric oxide, which re-coordinates to a stoichiometric $Ta_2O_5$ following the first hour of oxidation at room temperature. The binding energy of each oxide peak vs. the respective oxidation state is plotted above in Figure 5c/f as a means of determining the oxidation state of the a-Ta at time scales <ln(t)=8.25, since the binding energy and oxidation state are well aligned for both Nb (57) and $\alpha$-Ta (59), as reported in the literature. From this analysis, we conclude that $\alpha$-Ta forms a sub-stoichiometric $Ta_2O_{4.6}$ at the shortest time scales. The presence of this sub-stoichiometric oxide may offer an explanation for the reduced oxidation rate observed in Figure 3, and we speculate that the re-coordination of the oxide after ln(t)=8.25 yields an increase in oxidation rate due to higher field-assisted diffusivity of oxygen species in the stoichiometric oxide, though further analysis is needed to confirm this hypothesis.

**Conclusion:**

X-ray photoelectron spectroscopy and transmission electron microscopy reveal that the oxidation kinetics of superconducting Nb and $\alpha$-Ta differ at short time scales, with $\alpha$-Ta displaying a two-regime oxidation curve separated at ln(t)=8.25. Through TEM image calibration, we are able to calibrate the oxidation kinetic curves derived from XPS to surface oxide thickness. This calibration was verified through a comparison of near-zero-thickness at X = 0 with the Cabrera-Mott growth constant A extrapolated from the XPS data. Correlating our measurements of oxidation kinetics with device fabrication processes may prove valuable to the interfacial control of devices fabricated with these materials, as the surface oxide presence has been shown to directly lead to worsened coherence times. We observe that the tantalum oxide stoichiometry differs in the two oxidation regimes as determined by a shift in the relative binding energy of the $Ta^{+4.6}/Ta^{+5}$ oxide peak relative to the $Ta^{+0}$ metal peak, which could explain the shift in oxidation rate between the two regimes. Further investigations on crystallinity and diffusivity of oxygen species in these two regimes, with isotopic labeling, could investigate this phenomenon further and guide future work on surface oxide optimization. Broadly speaking, our method of data collection and analysis can provide a foundation for future studies of the oxidation of niobium and tantalum thin films to understand the structure of those interfaces and the oxide effect in superconducting quantum devices.

**Acknowledgements:**

We would like to acknowledge the support of the NY CREATES 300 mm fab team, and the support of the partners of the Center for Semiconductor Research. H. Frost would also like to acknowledge the support of TEL Technology Center, America, LLC. for his doctoral research.

**Conflicts of Interest:**

The authors have no conflicts to disclose.

**Data Availability:**

The data that support the findings of this study are available from the corresponding author upon reasonable request.

**Author Contributions:**

**Hunter J. Frost**: Conceptualization (equal); Data Curation (lead); Formal Analysis (lead); Investigation (lead); Methodology (lead); Validation (lead); Visualization (lead); Writing/Original Draft Preparation (lead); Writing/Review & Editing (equal). **Ekta Bhatia**: Conceptualization (equal); Data Curation (equal); Formal Analysis (equal); Investigation (equal); Methodology (equal); Validation (equal); Visualization (equal); Writing/Review & Editing (equal). **Zhihao Xiao**: Conceptualization (equal); Data Curation (equal); Formal Analysis (equal); Investigation (equal); Methodology (equal); Validation (equal); Visualization (equal); Writing/Review & Editing (equal). **Stephen Olson**: Formal Analysis (equal); Validation (equal); Visualization (equal); Writing/Review & Editing (equal). **Corbet Johnson**: Data Curation (supporting); Writing/Review & Editing (equal). **Kevin Musick**: Data Curation (supporting); Writing/Review & Editing (equal). **Thomas Murray**: Data Curation (supporting); Writing/Review & Editing (equal). **Christopher Borst**: Conceptualization (equal); Funding Acquisition (equal); Investigation (equal); Methodology (equal); Project Administration (equal); Resources (equal); Supervision (lead); Validation (lead); Visualization (lead); Writing/Review & Editing (equal). **Satyavolu Papa Rao**: Conceptualization (lead); Data Curation (equal); Formal Analysis (lead); Funding Acquisition (lead); Investigation (equal); Methodology (equal); Project Administration (lead); Resources (lead); Supervision (lead); Validation (lead); Visualization (lead); Writing/Review & Editing (equal).